\documentclass[prl,twocolumn,showpacs,preprintnumbers,amsmath,amssymb]{revtex4}

\usepackage{graphicx}
\usepackage{dcolumn}
\usepackage{bm}

\newcommand{\dd}{{\rm d}}

\newcommand{\rL}{{\mathrm{L}}}
\newcommand{\rR}{{\mathrm{R}}}

\newcommand{\mean}[1]{\overline{#1}}

\usepackage{amsmath}

\begin{document}

\title{An exactly solvable model of a highly efficient thermoelectric engine}


\author{Martin Horvat}
\email{martin.horvat@fmf.uni-lj.si}

\author{Toma\v z Prosen}
\email{tomaz.prosen@fmf.uni-lj.si}
\affiliation{Faculty of mathematics and physics, Department of Physics, University of Ljubljana, Jadranska 19, SI-1000 Ljubljana, Slovenia}

\author{Giulio Casati}
\email{giulio.casati@uninsubria.it}
\affiliation{Center for Nonlinear and Complex Systems, Via Vallegio, 11, 22100 Como Italia}

\date{\today}

\begin{abstract}
We propose a simple classical dynamical model of a thermoelectric (or
thermochemical) heat engine based on a pair of ideal gas containers
connected by two unequal scattering channels.
The model is solved analytically and it is shown that a suitable combination of parameters can be chosen
such that the engine operates at Carnot's efficiency.
\end{abstract}


\maketitle

In the frame of non-equilibrium thermodynamics an heat engine is a
machine generating work while exchanging  heat with two heat baths
at different temperatures $T_1$ and $T_2$.  The usual goal in a
construction of realistic heat engines is to increase the efficiency
as far as possible towards the theoretical upper limit $ \eta_{\rm
carnot} = 1-T_1/T_2$, assuming $T_2 > T_1$. We are here interested
in an engine without moving mechanical parts, i.e. which could
operate in a non-equilibrium {\em steady state}, such as for example
a thermoelectric or thermochemical couple. Such an engine - or a
refrigerator if the operation is reversed - would have immense
practical advantages over piston or compressor based engines for
obvious reasons, also due to possibilities of drastic
miniaturization \cite{mahan}.

Here we present an abstract model of a heat engine that can mimic
the essential features of a realistic heat engine based on the
thermoelectric effect \cite{degroot,linke,casati08} and which can be
treated and solved analytically. It is based purely on deterministic
classical dynamics and stochastic baths. The model is composed of
two thermochemical reservoirs of ideal gas of equal point particles
connected by two one-dimensional wires indexed by $i \in \{1,2\}$.
In the middle of each wire we place a deterministic and energy
conserving scatterer, which either reflect or transmit the particle
depending on its kinetic energy $\epsilon$. This behaviour is
completely described by the transmission function $\tau_i(\epsilon)
\in \{0,1\}$ of the $i-$th scatterer. We use units in which particle
mass $m$, particle charge $e$ and Boltzmann constant $k_{\rm B}$
equal $m=e=k_{\rm B}=1$. In this paper we show that in the steady state a non-vanishing
circular particle current exists only if the transmission
functions are energy dependent. Then we show that for a suitable combination of
parameters the engine operates in a reversible way with the Carnot's
efficiency.

\begin{figure}[!htb]
\centering
\includegraphics[width=8cm]{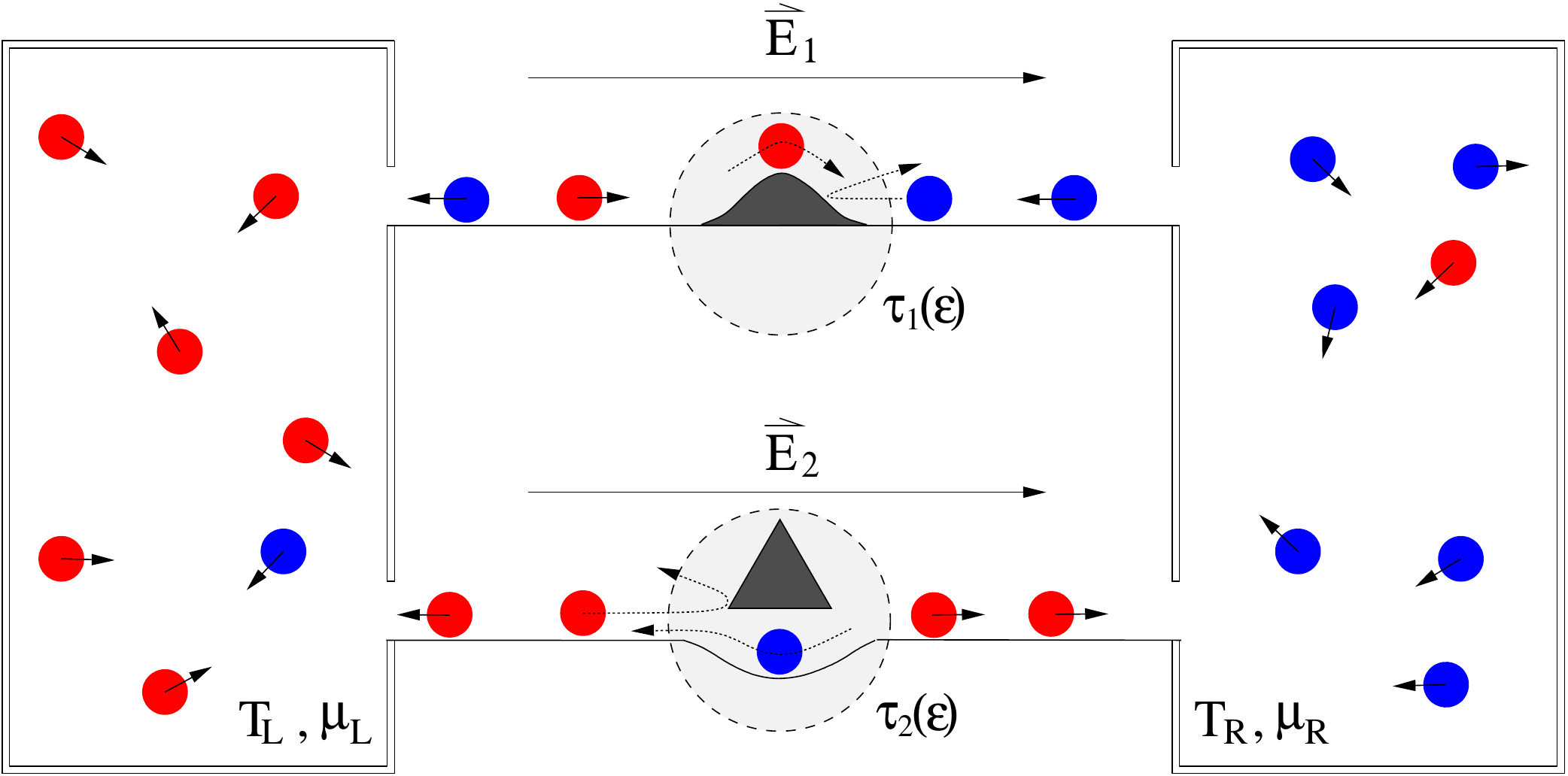}
\caption{Schematic figure of the heat engine. The possible deterministic scattering mechanisms are depicted symbolically.}
\label{pic:scheme}
\end{figure}

The scheme of the heat engine is shown in figure \ref{pic:scheme}.
In the wires we introduce {\em bias forces} $\vec{E}_i$ (say
electric fields), which can be described by bias voltages $U_i$ or
any other form of external potential energy which can be used to
extract useful work. In the stationary state, at some temperature
difference, there is a non-zero (circular) particle current in the
wires that, by climbing against the electric potential, can perform
useful work.
In the left (right) reservoir, the particles are at
chemical potentials $\mu_\rL$ ($\mu_\rR$) and temperature $T_\rL$
($T_\rR$) (here we assume $T_\rL > T_\rR$)   and are effused into
the wires with the injection rates $p_i \gamma_\rL$ ($p_i
\gamma_\rR$) into the first $i=1$ and the second $i=2$ channel,
respectively, where $p_i \in [0,1]$, $p_1+p_2=1$, represent the
relatives openings into the two channels. We note that both in the reservoirs and in
the channels the motion of particles is assumed to be (quasi) {\em one dimensional}, so we consider a single
component of the velocity. The injection rates are
connected to the chemical potentials $\mu_\nu$ and inverse
temperatures $\beta_\nu=1/T_\nu$ via the formula
\begin{equation}
  \mu_\nu \beta_\nu = \log(C\, \beta_\nu\gamma_\nu)\>,\qquad \nu\in\{\rL,\rR\}\>,
\end{equation}
where $C$ is a constant depending only on properties of particles
and on geometry of the reservoir opening \cite{casati08}. The
velocity $v$ of effused particles is distributed in each side
according to a canonical distribution
\begin{equation}
  P_\nu(v) = \beta_\nu v e^{-\frac{1}{2}\beta_\nu v^2} \theta(\sigma_\nu v) \>,
\end{equation}
where $\theta(v)=(1:v\ge 0; 0: {\rm otherwise})$ is the unit step function, and
$\sigma_\rL=1,\sigma_\rR=-1$. In the steady state, the particle currents $j_{\rho,i}$ within $i$-th wire and the heat currents
 \cite{degroot} exchanged with the $\nu$-side bath and $i$-th wire $j_{q,i}|_\nu$ are given by
\begin{align}
  j_{\rho,i}
  = p_i (\gamma_\rL t_{\rL,i} - \gamma_\rR t_{\rR,i}),\\
  j_{q,i}|_\rL
  = p_i( \gamma_\rL q_{\rL,i} - \gamma_\rR (q_{\rR,i} + t_{\rR,i} U_i)), \label{eq:heatL}\\
  j_{q,i}|_\rR
  = p_i( \gamma_\rL (q_{\rL,i} - t_{\rL,i} U_i) - \gamma_\rR q_{\rR,i})\label{eq:heatR}.
\end{align}
Here we have introduced the transmission
probability $t_{\nu,i}$  for a particle to transit from the $\nu$-side
to the other side over the $i$-th wire, and its average kinetic
energy $q_{\nu,i}$, explicitly defined in terms of the first two statistical moments of
the energy distribution of the effused particles transmitted through
the $i$th wire
\begin{eqnarray}
  \left(t_{\rL,i}, q_{\rm L,i}\right)
  &=& \beta_{\rL}\!\!\ \int_{\max\{0,U_i\}}^\infty\!\!\!\!\!\!\!\!\!\!\!\!\!\dd \epsilon\, e^{-\beta_{\rL} \epsilon} \tau_i\left(\epsilon - \frac{U_i}{2}\right) (1,\epsilon), \\
    \left(t_{\rR,i}, q_{\rm R,i}\right)
  &=& \beta_{\rR}\!\!\ \int_{\max\{0,-U_i\}}^\infty\!\!\!\!\!\!\!\!\!\!\!\!\!\!\!\!\!\!\dd \epsilon\, e^{-\beta_{\rR} \epsilon} \tau_i\left(\epsilon + \frac{U_i}{2}\right) (1,\epsilon).
\end{eqnarray}
The terms $U_i/2$ in the arguments of transmission functions
$\tau_i$ imply the assumption of linear potential and scatterers
being in the middle of each wire. However different assumptions (say
of bias potential steps at the left/right of each scatterer) could
be treated straightforwardly.
By imposing the condition of stationarity  $j_{\rho, 1} + j_{\rho,2}
= 0$, we obtain, from eq. (3),  the injection rates $\gamma_\rL$ and
$\gamma_\rR$
\begin{equation}
  1-\frac{\gamma_\rL}{\overline{\gamma}} =
  \frac{\gamma_\rR}{\overline{\gamma}}-1 =
  \frac{p_1(t_{\rL,1}-t_{\rR,1}) + p_2(t_{\rL,2}-t_{\rR,2})}
         {p_1(t_{\rL,1}+t_{\rR,1}) + p_2(t_{\rL,2}+t_{\rR,2})}\>.
\end{equation}
where $\overline{\gamma} = (\gamma_\rL + \gamma_\rR)/2$ \cite{note}.
The resulting particle current in the wires
\begin{equation}
  j_{\rho,1} = 2 \overline{\gamma} p_1 p_2
  \frac{t_{\rL,1} t_{\rR,2} -t_{\rL,2} t_{\rR,1}}
       {p_1(t_{\rL,1}+t_{\rR,1}) + p_2(t_{\rL,2}+t_{\rR,2})}\>,
\end{equation}
determines the working power $P =  j_{\rho,1} (U_1 - U_2)$ while the
{\em ingoing} heat flux is equal to
\begin{align}
  Q &= j_{q,1}|_\rL + j_{q,2}|_\rL\>,& \nonumber\\
    &=-\frac{2 \overline{\gamma} \sum_{i,j=1}^2 p_i p_j t_{\rL,i} t_{\rR,j} (U_i - S_{\rL, i} + S_{\rR,j}) }
        {p_1(t_{\rL,1}+t_{\rR,1}) + p_2(t_{\rL,2}+t_{\rR,2})}\>,&
\end{align}
where we have introduced the ratio $S_{\nu,i} = q_{\nu,i}/t_{\nu,i}$
(which is shown below to be connected to the Seebeck coefficient).
The efficiency of the heat engine is then defined as $\eta = P/|Q|$.
Notice that the particle current $j_{\rho,1}$, and so also the power $P$, is
proportional to the determinant of the matrix of transmission
coefficients $D=\det\{t_{\nu,i}\} = t_{\rL,1} t_{\rR,2} - t_{\rL,2}
t_{\rR,1}$. The optimal performance of the heat engine for a given
configuration of temperatures and scatterers, is obtained by finding
appropriate fields $U_1$ and $U_2$ which maximize $\eta(U_1,U_2)$.
This can be done, in general, only numerically since it requires a
solution of coupled transcendental equations.

Note the following important observation: $D=0$, and hence the
currents vanish, despite non-vanishing temperature difference, if
the scatterers are energy independent $\tau_i(\epsilon) \equiv {\rm
const}$. This fact is a simple
consequence of time-reversal properties of individual deterministic
trajectories which connect the two baths and remains valid for scattering channels in
higher dimension (e.g. like in Ref.~\cite{casati08}).

\begin{figure}[!htb]
\centering
\includegraphics[width=7.5cm]{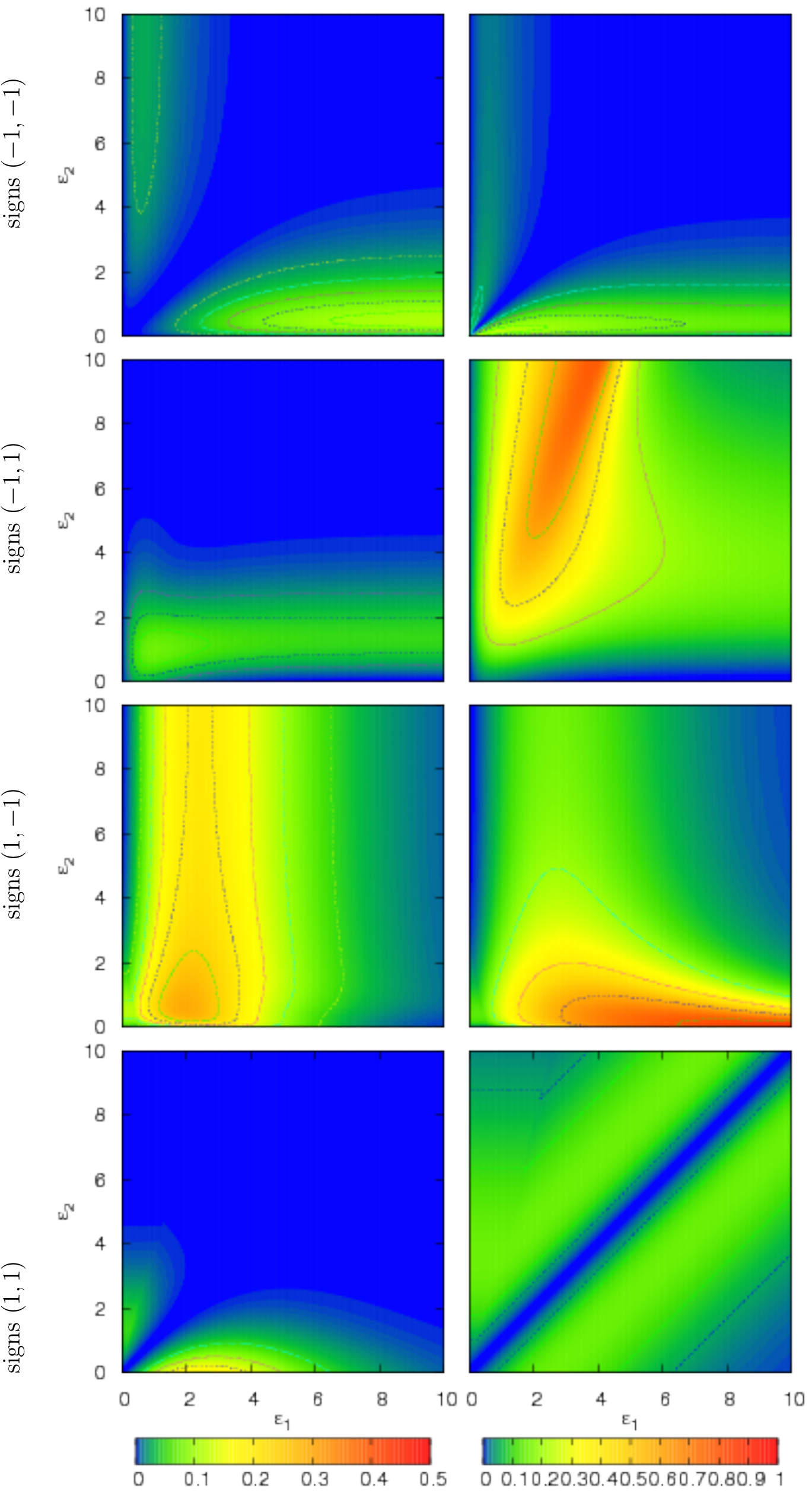}
\hbox to 8cm{\hfil(a)\hfil(b)\hfil}
\caption{The relative power $P^*/\eta_{\rm carnot}^2$ (a) and the relative efficiency
$\eta^*/\eta_{\rm carnot}$ (b) as functions of the energy thresholds $\epsilon_i$
for different sign configurations $(s_1,s_2)$ (indicated on the left) at unit
mean injection rate $\mean{\gamma}=1$, equal channel openings $p_i=1/2$,
 and bath temperatures $T_\rL=2$ and $T_\rR=0.5$.}
\label{pic:nonlin}
\end{figure}

Our ideas are demonstrated in a heat engine with bias voltage only
in the first wire ($U_2=0$) and for the simplest nontrivial,
step-like, transmission functions
\begin{equation}
\tau_i(\epsilon)= \theta(s_i (\epsilon - \epsilon_i)).
\label{eq:tf}
\end{equation}
The direction of the steps, at the energy thresholds $\epsilon_i$,
is determined by the signs $s_i \in\{1,-1\}$. Simple mechanical
realizations for both signs $s_i \in\{1,-1\}$ are schematically
depicted in fig.~\ref{pic:scheme}. We numerically determine the
potential $U_1$ that maximizes the efficiency for a
given configuration of scatterers. The optimal efficiency
$\eta^*=\max_{U_1}\eta$ and the corresponding power $P^*$ are shown
in figure \ref{pic:nonlin}.
In the case $s_1=s_2=-1$ the scatterers transmit only slow enough particles. The regions of high power and high efficiency overlap and are
 positioned almost symmetrically near the axes. The exact symmetry is broken because the electric field is only applied to the first wire.
  The cases $s_1=-s_2=-1$ and $s_1=-s_2=1$ describe a similar situation, where one scatterer transmit the fast particles and the other scatterer
  transmit slower ones. This case is the most efficient and  $\eta$ here may nearly approach $\eta_{\rm carnot}$. However, as expected,
  the regions of high efficiency and high power only slightly overlap. 
  In the last case $s_1=s_2=1$ the scatterers only transmit fast enough particles. The region of high efficiency is located parallel to
   the line $\epsilon_1=\epsilon_2$. The highest power is obtained for energy steps matching the baths temperatures
  $\epsilon_1=T_\rL$, $\epsilon_2=T_\rR$. A detailed analysis of the relaxation process shows that the convergence time to
   the non-equilibrium steady state is {\em strictly finite} for non-vanishing bias potentials, and is given by $\tau^* \sim 1/\min\{|U_1|,|U_2|\}$.

In the following we show that the results drastically simplify in
the {\em linear regime} of small relative temperature difference. In
this regime the meaningful bias potentials are also small and we may
approximate the exact particle and heat fluxes in the wires with
their linear expansions in the temperature difference $T_\rL-T_\rR$,
injection rate difference $\delta\gamma=\gamma_\rR-\gamma_\rL$, and
potentials $U_i$. Expressing
$\delta\gamma$ with the chemical potential difference
$\delta\mu=\mu_\rR-\mu_\rL$, setting $\mu_\rR+\mu_\rL = 0$, we write
the particle and heat fluxes in the linear response limit as
\begin{align}
j_{\rho,i}
= p_i \bar{\gamma} (-\beta g_i (U_i + \delta \mu) + h_i\delta \beta)\>, \\
j_{q,i}
= p_i \bar{\gamma} (-\beta h_i (U_i + \delta \mu) + k_i \delta \beta)\>,
\end{align}
where $\beta = (\beta_\rL + \beta_\rR)/2$ and $\delta\beta =\beta_\rR - \beta_\rL$.
The expansion coefficients $g_i$, $h_i$ and $k_i$ are
statistical moments of a canonical energy distribution of particles
that are transmitted over the $i$-th wire
\begin{equation}
  (g_i,h_i ,k_i)
  = \beta \int_{0}^\infty \dd \epsilon\,e^{-\beta\epsilon} \tau_i(\epsilon) (1,\epsilon,\epsilon^2)
  \label{eq:moments}
\end{equation}
and depend only on the transmission function and temperature. Notice
that in the linear response limit the heat fluxes at the left and
right side are equal in contrast to the general (non-linear) case
(\ref{eq:heatL}), (\ref{eq:heatR}). The coefficients $g_i$ and $h_i$
represent the average transmission probability of  particles across
the $i$-th wire and their average energy at zero bias fields.
Instead of $h_i$ and $k_i$ it is more convenient to work with the
average energy per particle $S_i = h_i/g_i$ and the coefficient $K_i
= k_i -g_i S_i^2$. Note that $\beta p_i \bar{\gamma} g_i$, $\beta^2
p_i\bar{\gamma} K_i$ and $\beta S_i$ can be interpreted as the {\em
particle conductance}, the {\em heat conductance} and the {\em
Seebeck coefficient}, respectively. By imposing the stationarity
condition $j_{\rho,1}+j_{\rho,2}=0$ we obtain the difference of the
chemical potentials between baths:
\begin{equation}
  \delta \mu = -
  \frac{\xi (p_1 g_1 S_1 +  p_2 g_2 S_2) + p_1 g_1 U_1 + p_2 g_2 U_2}
         {p_1 g_1 + p_2 g_2}\>,
\end{equation}
where $\xi = -\delta \beta/\beta$ is the relative temperature
difference which is related to the Carnot efficiency $\eta_{\rm
carnot} = |\xi|$.

\begin{figure}[!htb]
\centering
\includegraphics[width=7.5cm]{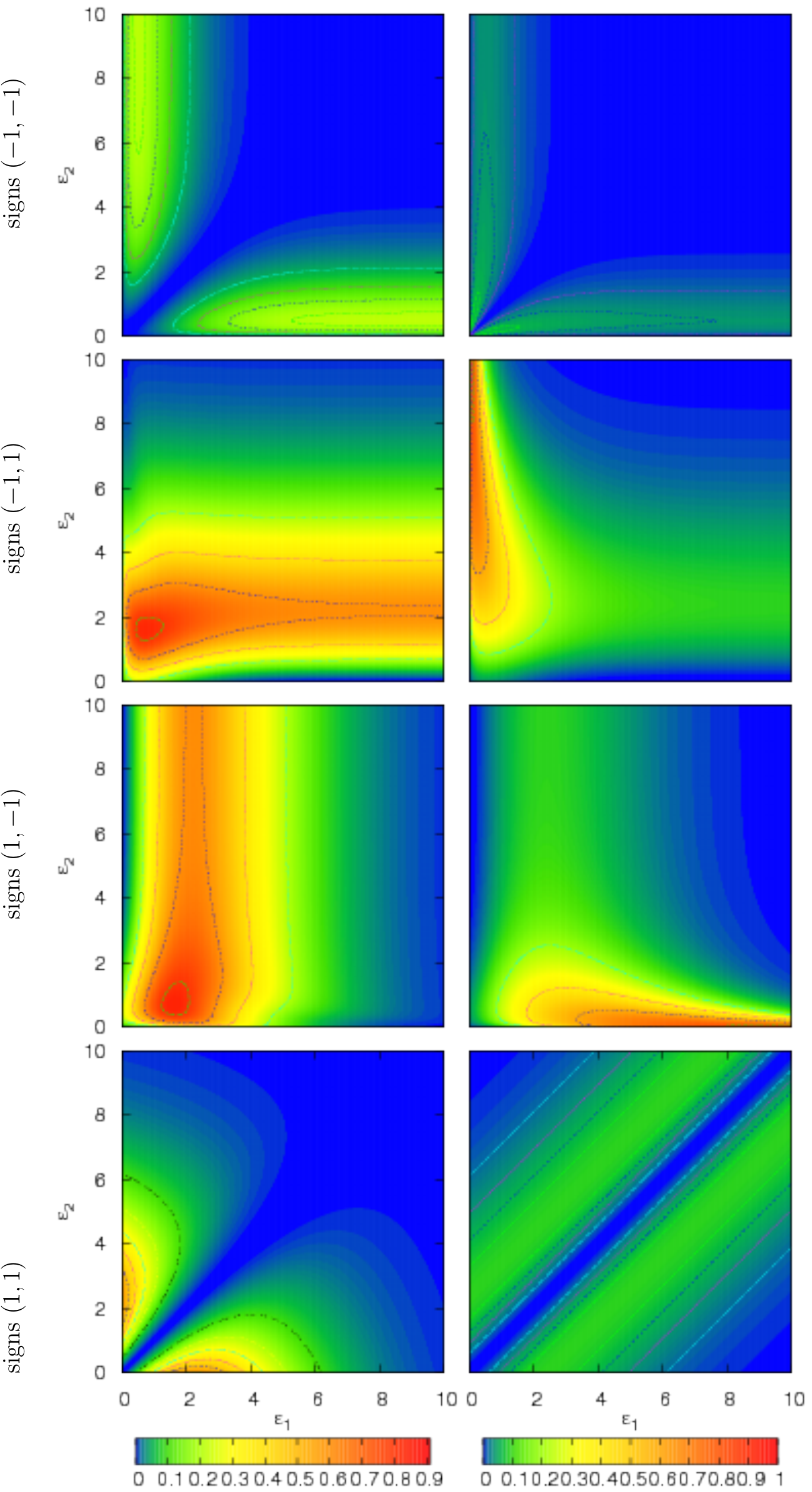}
\hbox to 8cm{\hfil(a)\hfil(b)\hfil}
\caption{The density plot of the relative power $10P^*/\eta_{\rm carnot}^2$ (a) and the relative efficiency $\eta^*/\eta_{\rm carnot}$ (b) in the linear response regime, at the bath temperatures $T_\rL=1.01$ and $T_\rR=0.99$. (Other details the same as in figure \ref{pic:nonlin})}
\label{pic:lin}
\end{figure}

Let us now introduce the auxiliary quantities: difference of
energies per particle in the two wires $\Delta S = S_2 - S_1$,
difference of potentials $\delta U = U_2 - U_1$,
transmission probability through both wires $G = ((p_1 g_1)^{-1} +
(p_2 g_2)^{-1})^{-1}$, and the figure of merit of the heat engine
efficiency \cite{jiao}
\begin{equation}
  y = \frac{G (\Delta S)^2 } {p_1 K_1 + p_2 K_2}> 0 \>.
\end{equation}
We can now write the particle current in the first wire $j_{\rho,1}$
and the ingoing heat flux $Q= j_{q,1}+ j_{q,2} $ elegantly as
\begin{align}
 j_{\rho,1} = \beta\gamma G (-\delta U + \xi \Delta S)\>,\\
 Q = \beta\gamma G (\Delta S\, \delta U - \xi (\Delta S)^2(1+1/y))\>,
\end{align}
whereby the power and the efficiency of the heat engine are $P=j_{\rho, 1} \delta U$ and $\eta= P/|Q|$, respectively. Notice that all expressions just depend on the potential difference $\delta U$. In the linear response regime the potential that maximises the efficiency can be found analytically by solving the equation $\partial \eta /\partial (\delta U)=0$. The explicit solutions are
\begin{align}
  \delta U^* = \xi \Delta S (1 - (\sqrt{1+y}-1)/y) \>,
  \label{eq:optlin_pot}\\
  j_{\rho,1}^* = \xi \bar{\gamma} \beta  G \Delta S (\sqrt{1+y}-1)/y\>,
  \label{eq:optlin_current}\\
  Q^* = -\xi \bar{\gamma} \beta G (\Delta S)^2 \sqrt{1 + y}/y\>,
\end{align}
yielding the optimal efficiency (equivalent to Eq. (14) of \cite{jiao})
\begin{equation}
  \eta^* = \frac{P^*}{|Q^*|}
         = \eta_{\rm carnot} \left(1 + \frac{2}{y}\left(1 - \sqrt{1+y}\right)\right)\>,
  \label{eq:opteff}
\end{equation}
with the corresponding power $P^* = j_{\rho,1}^* \delta U^*$. The
relative optimal efficiency $\eta^*/\eta_{\rm carnot}$ , as
expected, depends only on $y$ and is monotonic in the latter.
Therefore it is meaningful to treat $y$ as the figure of merit of heat engine
efficiency. We note that if the transmission functions are given by
(\ref{eq:tf}), the results (\ref{eq:optlin_pot}-\ref{eq:opteff}) are
{\em explicit} as all the expressions are explicit rational
functions of the moments of the Laplace transform of the
transmission functions (\ref{eq:moments}), which in turn are simple
algebraic functions of $\beta,\epsilon_i$ and $\exp(-\beta
\epsilon_i)$. From the equations (\ref{eq:optlin_pot}) and
(\ref{eq:optlin_current}) we can recognize that $P^*\propto
\eta_{\rm carnot}^2$, and consequently, in the linear response regime the power-output is
rather small. The optimal efficiency and the corresponding power as
function of $\epsilon_1$ and $\epsilon_2$ in linear response regime
are shown in figure \ref{pic:lin}.
They are quite similar to those obtained in the non-linear regime
shown in figure \ref{pic:nonlin}. The important difference between
nonlinear and linear regime results is that the latter only depends
on the difference $\delta U$ of bias potentials  and can be made
temperature independent by expressing the energy steps in  the
transmission function $\tau_i(\epsilon)$ with the parameters $r_i =
\beta\epsilon_i$ . Consequently, the power $P^*$ and the efficiency
$\eta^*$ as function of $(\epsilon_1,\epsilon_2)$ in the cases
$s_1=-s_2=1$ and $s_1=-s_2=-1$ are exactly symmetric w.r.t. exchange
of the parameters $\epsilon_i$.  These cases are also the most
efficient. The region of high efficiency is squeezed towards the
$\epsilon_2$ or $\epsilon_1$ axes in the cases $s_1=-s_2=1$ and
$s_1=-s_2=-1$, respectively.

We have performed exact analytical calculations of expressions (\ref{eq:optlin_pot}-\ref{eq:opteff}) for the case of equal channel
openings $p_i=1/2$. In the cases $s_1=s_2=1$ and $s_1=s_2=-1$ the maximal efficiency $\eta_{\rm max}=\max_{r_1,r_2>0} \eta^*$ is reached at
finite $(r_1,r_2)$ and is equal to $\eta_{\rm max} \doteq 0.066\, \eta_{\rm carnot}$ and $\eta_{\rm max} \doteq 0.091\, \eta_{\rm carnot}$,
respectively. However, in the case $s_1=-s_2=1$ (and similarly for $s_1=-s_2=-1$) we can reach the Carnot efficiency in the
limit $r_1\to \infty$ following the curve $r_2^2 \asymp 2 \exp(-r_1)$ along which the efficiency algebraically increases as
$
  \eta^* \asymp \eta_{\rm carnot} (1-2 r_1^{-1})$,
and the power exponentially decreases as
$
  P^* \asymp 2 \eta_{\rm carnot}^2 \bar{\gamma}\beta^{-1} e^{-r_1} \left(r_1-1 + \frac{5}{2 r_1}\right).
$

In conclusions, we have proposed a simple exactly solvable classical-mechanical model of thermoelectric (or better to say, thermochemical) heat engine. We presented closed form solutions for the steady state of the engine in linear and non-linear regimes. A variable thermodynamic efficiency has been found, as a function of the system's parameters, which can become arbitrary close to Carnot's in an appropriate regime.

Finally we would like to draw the reader attention to the following
point: it is possible to argue that our model is quite abstract in
nature and therefore far from possible realistic implementations. We
think on the contrary that this is the main advantage of our
approach. After more than 50 years during which thermoelectric
efficiency did not substantially increase we propose here to take a
completely opposite point of view. Starting from fundamental
microscopic equations and considering the most general schematized
framework, we hope to understand the basic dynamical mechanisms which can lead
to an increase of thermoelectric efficiency. In this spirit, the
model discussed here is a step in this direction. In addition our
model should be relevant for a theoretical description of nanoscopic
heat engines, for example  a pair of thermoelectrically coupled
quantum dots. However, our model would be a good approximation to the real
system only in a rather restricted situation of (i) non-interacting charge carriers, (ii) negligible phonon contributions to heat transport, (iii) coherence length longer than wires, which (iv) should be quasi one-dimensional.


\begin{thebibliography}{99}
\bibitem{note}  We note that the injection rates are simply related
to the density of particles $\rho_\nu$ and temperatures $T_\nu$ in the baths, as  $\gamma_\nu \propto \rho_\nu \sqrt{T}_\nu$,
with the proportionality factor depending only on the geometry of the channel (wire)
openings \cite{casati08}. However in order to make resulting expressions simpler we shall express them in terms of $\bar{\gamma}$.
\end{thebibliography}
\end{document}